\input harvmac

\input epsf

\newcount\figno
\figno=0 
\def\fig#1#2#3{
\par\begingroup\parindent=0pt\leftskip=1cm\rightskip=1cm\parindent=0pt
\baselineskip=11pt
\global\advance\figno by 1
\midinsert
\epsfxsize=#3
\centerline{\epsfbox{#2}}
\vskip 12pt
{\bf Fig.\ \the\figno: } #1\par
\endinsert\endgroup\par
}
\def\figlabel#1{\xdef#1{\the\figno}}

\lref\BMN{
D.~Berenstein, J.~M.~Maldacena and H.~Nastase,
``Strings in flat space and pp waves from N = 4 super Yang Mills,''
JHEP {\bf 0204}, 013 (2002)
[arXiv:hep-th/0202021].
%%CITATION = HEP-TH 0202021;%%
}

\lref\BlauNE{
M.~Blau, J.~Figueroa-O'Farrill, C.~Hull and G.~Papadopoulos,
``A new maximally supersymmetric background of IIB superstring theory,''
JHEP {\bf 0201}, 047 (2002)
[arXiv:hep-th/0110242].
}

\lref\gkp{
S.~S.~Gubser, I.~R.~Klebanov and A.~W.~Peet,
``Entropy and Temperature of Black 3-Branes,''
Phys.\ Rev.\ D {\bf 54}, 3915 (1996)
[arXiv:hep-th/9602135].
%%CITATION = HEP-TH 9602135;%%
}

\lref\GKT{
S.~S.~Gubser, I.~R.~Klebanov and A.~A.~Tseytlin,
``Coupling constant dependence in the thermodynamics of N = 4
supersymmetric Yang-Mills theory,''
Nucl.\ Phys.\ B {\bf 534}, 202 (1998)
[arXiv:hep-th/9805156].
%%CITATION = HEP-TH 9805156;%%
}

\lref\AMP{A.M. Polyakov, ``Gauge fields and space-time,''
[arXiv:hep-th/0110196].
%%CITATION = HEP-TH 0110196;%%
}

\lref\BerensteinSA{
D.~Berenstein and H.~Nastase,
``On lightcone string field theory from super Yang-Mills and holography,''
arXiv:hep-th/0205048.
%%CITATION = HEP-TH 0205048;%%
}

\lref\foto{
A.~Fotopoulos and T.~R.~Taylor,
``Comment on two-loop free energy in N = 4 supersymmetric Yang-Mills
theory at finite temperature,''
Phys.\ Rev.\ D {\bf 59}, 061701 (1999)
[arXiv:hep-th/9811224].
%%CITATION = HEP-TH 9811224;%%
}

\lref\vm{
M.~A.~Vazquez-Mozo,
``A note on supersymmetric Yang-Mills thermodynamics,''
Phys.\ Rev.\ D {\bf 60}, 106010 (1999)
[arXiv:hep-th/9905030];
%%CITATION = HEP-TH 9905030;%%
C.~j.~Kim and S.~J.~Rey,
``Thermodynamics of large-N super Yang-Mills theory and AdS/CFT
correspondence,''
Nucl.\ Phys.\ B {\bf 564}, 430 (2000)
[arXiv:hep-th/9905205];
%%CITATION = HEP-TH 9905205;%%
A.~Nieto and M.~H.~Tytgat,
``Effective field theory approach to N = 4 supersymmetric Yang-Mills at
finite temperature,''
arXiv:hep-th/9906147.
%%CITATION = HEP-TH 9906147;%%
}

\lref\zanon{
A.~Santambrogio and D.~Zanon,
``Exact anomalous dimensions of N = 4 Yang-Mills operators with large R
charge,''
arXiv:hep-th/0206079.
%%CITATION = HEP-TH 0206079;%%
}

\lref\KiemXN{
Y.~j.~Kiem, Y.~b.~Kim, S.~m.~Lee and J.~m.~Park,
``pp-wave / Yang-Mills correspondence: An explicit check,''
arXiv:hep-th/0205279.
%%CITATION = HEP-TH 0205279;%%
}

\lref\LeeRM{
P.~Lee, S.~Moriyama and J.~w.~Park,
``Cubic interactions in pp-wave light cone string field theory,''
arXiv:hep-th/0206065.
%%CITATION = HEP-TH 0206065;%%
}

\lref\grossetal{
D.~J.~Gross, A.~Mikhailov and R.~Roiban,
``Operators with large R charge in N = 4 Yang-Mills theory,''
arXiv:hep-th/0205066.
%%CITATION = HEP-TH 0205066;%%
}

\lref\seven{
N.~R.~Constable, D.~Z.~Freedman, M.~Headrick, S.~Minwalla, L.~Motl,
A.~Postnikov and W.~Skiba,
``PP-wave string interactions from perturbative Yang-Mills theory,''
arXiv:hep-th/0205089.
%%CITATION = HEP-TH 0205089;%%
}

\lref\gkppp{
S.~S.~Gubser, I.~R.~Klebanov and A.~M.~Polyakov,
``A semi-classical limit of the gauge/string correspondence,''
arXiv:hep-th/0204051;
%%CITATION = HEP-TH 0204051;%%
S.~Frolov and A.~A.~Tseytlin,
``Semiclassical quantization of rotating superstring in AdS(5) x S**5,''
JHEP {\bf 0206}, 007 (2002)
[arXiv:hep-th/0204226].
%%CITATION = HEP-TH 0204226;%%
}

\lref\gs{
M.~B.~Green and J.~H.~Schwarz,
``Superstring Interactions,''
Nucl.\ Phys.\ B {\bf 218}, 43 (1983).
%%CITATION = NUPHA,B218,43;%%
}

\lref\huang{
M.~x.~Huang,
``Three point functions of N = 4 super Yang Mills from light cone string
field theory in pp-wave,''
arXiv:hep-th/0205311.
%%CITATION = HEP-TH 0205311;%%
}

\lref\mn{
M.~Spradlin and A.~Volovich,
``Superstring interactions in a pp-wave background,''
arXiv:hep-th/0204146.
%%CITATION = HEP-TH 0204146;%%
}

\lref\mntwo{
M.~Spradlin and A.~Volovich,
``Superstring interactions in a pp-wave background. II,''
arXiv:hep-th/0206073.
%%CITATION = HEP-TH 0206073;%%
}

\lref\MetsaevBJ{
R.~R.~Metsaev,
``Type IIB Green-Schwarz superstring in plane wave Ramond-Ramond  background,''
Nucl.\ Phys.\ B {\bf 625}, 70 (2002)
[arXiv:hep-th/0112044].
%%CITATION = HEP-TH 0112044;%%
}

\lref\MetsaevRE{
R.~R.~Metsaev and A.~A.~Tseytlin,
``Exactly solvable model of superstring in plane wave
Ramond-Ramond  background,''
Phys.\ Rev.\ D {\bf 65}, 126004 (2002)
[arXiv:hep-th/0202109].
%%CITATION = HEP-TH 0202109;%%
}

\lref\GSB{
M.~B.~Green, J.~H.~Schwarz and L.~Brink,
``Superfield Theory Of Type II Superstrings,''
Nucl.\ Phys.\ B {\bf 219}, 437 (1983).
%%CITATION = NUPHA,B219,437;%%
}

\lref\KristjansenBB{
C.~Kristjansen, J.~Plefka, G.~W.~Semenoff and M.~Staudacher,
``A new double-scaling limit of N = 4 super Yang-Mills theory and PP-wave
strings,''
arXiv:hep-th/0205033.
%%CITATION = HEP-TH 0205033;%%
}

\lref\ChuPD{
C.~S.~Chu, V.~V.~Khoze and G.~Travaglini,
``Three-point functions in N = 4 Yang-Mills theory and pp-waves,''
JHEP {\bf 0206}, 011 (2002)
[arXiv:hep-th/0206005].
%%CITATION = HEP-TH 0206005;%%
}

\lref\Herman{
H.~Verlinde,
``Bits, Matrices and 1/N,''
arXiv:hep-th/0206059.
}

\Title
{\vbox{
 \baselineskip12pt
\hbox{hep-th/0206221}
\hbox{PUTP-2042}
\hbox{HUTP-02/A027}
}}
 {\vbox{
 \centerline{New Effects in Gauge Theory}
\centerline{}
 \centerline{from pp-wave Superstrings}
 }}

\centerline{
Igor R. Klebanov${}^{1}$,
Marcus Spradlin${}^{1}$ and Anastasia Volovich${}^{2}$
}

\bigskip
\centerline{${}^{1}$~Department of Physics}
\centerline{Princeton University}
\centerline{Princeton, NJ 08544}
\centerline{\tt klebanov, spradlin@feynman.princeton.edu}
\centerline{}
\centerline{${}^{2}$~Department of Physics}
\centerline{Harvard University}
\centerline{Cambridge, MA 02138}
\centerline{\tt nastya@gauss.harvard.edu}

\vskip .3in
\centerline{\bf Abstract}
It has recently been observed that IIB string theory in
the pp-wave background can be used to calculate certain
quantities, such as the dimensions of BMN operators, as
exact functions of the effective coupling $\lambda' = \lambda/J^2$.
These functions
interpolate smoothly between the weak and strong effective
coupling regimes of ${\cal{N}}=4$ SYM theory at large R charge $J$.
In this paper we use the pp-wave superstring field theory 
of hep-th/0204146 to study
more complicated observables.
The expansion of the three-string interaction vertex suggests
more complicated interpolating functions
which in general give rise to fractional powers
of $\lambda'$ in physical observables at weak effective coupling.
\smallskip

\Date{}

\listtoc
\writetoc

\newsec{Introduction}

Recently the exact solvability
\refs{\MetsaevBJ, \MetsaevRE}
of type IIB string theory in the pp-wave
background \BlauNE\ has
been used to understand the AdS/CFT correspondence
in the limit of large R charge \BMN. It was discovered that string theory
makes new exact statements about the ${\cal N}=4$
SYM theory that may be checked in perturbation theory.
The simplest such prediction concerns the dimensions of the BMN
operators of R charge $J$ \BMN:
\eqn\bmndim{
\Delta - J =\sum_{n=-\infty}^\infty N_n \sqrt {1 + \lambda' n^2
}
}
This formula shows that, for large $J$, these dimensions are functions
of $\lambda'= \lambda/J^2$ where $\lambda$ is the `t Hooft
coupling.\foot{This result was rederived in \gkppp\ (following
earlier suggestions in \AMP)
 via semiclassical analysis of
the $AdS_5\times S^5$ sigma model, valid for large $\lambda$ and large
$J$.} 
Therefore, even though the stringy derivation of this formula assumes
that $\lambda$ is large, 
the effective coupling $\lambda'$ is a parameter that may assume
arbitrary
values. The interpolating formula \bmndim\ is remarkable: not only does
it have the correct strong and weak coupling limits, but it constitutes
a string theoretic prediction for perturbative gauge theory, which has
recently been checked successfully \refs{\grossetal,\zanon}.
Further interesting gauge theory results for correlators of the BMN
operators were obtained in \refs{\KristjansenBB  \BerensteinSA-\seven}.
In order to compare these results with string theory, it
is important to develop a string theoretic approach
to observables
more complicated than the operator dimensions; for example, the
3-point functions of the BMN operators.
Since the RR-charged pp-wave background is solvable in the
light-cone gauge, it is appropriate to use the techniques of light-cone
superstring field theory \refs{\gs, \GSB}.
Extension of this formalism from flat space to the pp-wave
background was presented in \mn\ and further explored in
\refs{\mntwo \huang  \ChuPD \KiemXN-\LeeRM}.
In this paper we present some additional
calculations which shed new light on the $\lambda'$ dependence
of various observables.

Consider, for example, 
3-point functions of the BMN operators with large R charge.
While the position dependence is fixed in terms of the operator
dimensions by the conformal invariance, the normalization
$C_{ijk}$ is an interesting observable.
If we restrict ourselves
to the planar limit, $C_{ijk}$ may depend on the `t Hooft
coupling $\lambda$ and the R charges $J_i$ through
combinations $\lambda'= \lambda/J_1^2$ and $J_2/J_1$,
and there is no a priori reason to
believe that the dependence  is particularly simple.
We will argue that in this case the interpolating function may be
far more complex than \bmndim\ and will present some evidence for this.

An analogy we have in mind is to another non-BPS observable: the
free energy of the ${\cal N}=4 $ SYM theory at a finite temperature $T$.
On general field theoretic
grounds we expect that in the `t Hooft large $N$
limit the entropy is given by
\eqn\free{
F/V= -{ \pi^2\over 6} N^2 f(\lambda) T^4
\ .
}
The AdS/CFT correspondence predicts the following behavior of $f$
for large $\lambda$ \refs{\gkp,\GKT}:
\eqn\strong{
f(\lambda) = {3\over 4} +
{45\over 32} \zeta(3) \lambda^{-3/2}  + \ldots
\ .
}
On the other hand, perturbative field theory gives the following
small $\lambda$ behavior \refs{\foto,\vm}:
\eqn\weak{
f(\lambda) = 1 - {3\over 2\pi^2} \lambda
+{3+\sqrt 2\over \pi^3} \lambda^{3/2}
  + \ldots
\ .
}
Calculation of the full interpolating function is an interesting challenge
which seems to be beyond the scope of presently available methods:
supergravity methods are not sufficient for studying small $\lambda$
while full string theoretic methods have not been developed far enough.
The expansions \strong\ and \weak\
 indicate, however, that the interpolating function is far more
complicated than in \bmndim. For instance, at small $\lambda$ we observe
the appearance of a term of order $\lambda^{3/2}$ 
\vm\ which is due to a
resummation of diagrams with insertions of
the thermal mass induced at one loop, $m^2\sim \lambda T^2$.
This non-analytic term is an infrared effect: it follows from the
fact that the free energy depends on the mass as $F/V\sim m^3 T$.
In this paper we will see hints of similar effects 
in the pp-wave light-cone string field theory.
Luckily, in this case methods are
available for studying the string field theory at small $\lambda'$ (or,
equivalently\foot{In this paper we
use $\mu$ as shorthand for the dimensionless variable
$1/\sqrt{\lambda'}$.  This is a departure from the
more conventional relation $\lambda' = {1 \over (\mu p^+ \alpha')^2}$,
where $p^+$ is the largest light-cone momentum involved
in the process of interest.}, at large $\mu$).
We turn to this analysis in the
next section.

\newsec{The Light-Cone String Vertex at Large $\mu$}

The three string splitting-joining interaction in the pp-wave
background has been worked out in \mn.
The interaction consists of a delta-functional overlap
which expresses continuity of the string worldsheet,
and an operator required by supersymmetry
which is inserted at the point where the string splits \mntwo.
In this paper we focus on the overlap, which we express
as a state in the three-string Hilbert space of the form
\eqn\vertex{
|V\rangle = \exp \left[ {1 \over 2} \sum_{r,s=1}^3
\sum_{m,n=-\infty}^\infty a^{I \dagger}_{m(r)}
\overline{N}^{(rs)}_{mn} a^{J \dagger}_{n(s)} \delta_{IJ} \right] |0\rangle.
}
Like the dimensions \bmndim, the Neumann coefficients
$\overline{N}^{(rs)}_{mn}$
are smooth
functions of $\lambda'$ which interpolate between the flat space
expressions of \gs\ at $\lambda' = \infty$ and the very simple
expressions of \refs{\mntwo,\huang} at $\lambda' = 0$.
They encode a wealth of information about the
interacting gauge theory, but
unlike \bmndim\
they are highly nontrivial
functions of $\lambda'$ which have not been computed
explicitly.
In this section we report some progress in this direction.
We highlight the difficulty of calculating even
${\cal{O}}(\lambda')$
effects, and point out the existence of non-analytic terms
involving half-integer powers of $\lambda'$ as
well as $e^{-1/\sqrt{\lambda'}}$.

\subsec{The Matrix $\Gamma_+$}

The difficulty in obtaining explicit formulas for
the Neumann coefficients starts with the problem of
inverting a certain infinite dimensional matrix
$\Gamma_+$.  In appendix A we define this matrix and
evaluate its components explicitly.  It can be
expressed as
\eqn\gp{
\Gamma_+ = \Gamma_0 - H
}
where $\Gamma_0$ is diagonal,
\eqn\godef{
\left[ \Gamma_0 \right]_{mn} = 2 {\sqrt{m^2 + \mu^2}
\over m} \delta_{mn},
}
(for positive integers $m,n$)
and the matrix elements of $H$ are
\eqn\hdef{
H_{mn} = {8 \over \mu^2 \pi^2} (-1)^{m+n}
\sqrt{m n} \sin(\pi m y) \sin(\pi n y)
\int_1^\infty dz { F(z) \sqrt{z^2 - 1}
\over (z^2 + m^2/\mu^2)(z^2 + n^2/\mu^2)},
}
where $y = p^+_{(1)}/p^+_{(3)}$ lies in
the range $0<y<1$ and
\eqn\fdef{
F(z) = \ha \left[
\coth(\pi \mu y z) + \coth(\pi \mu (1-y) z)\right].
}
Note that $H$ has a finite limit as $\mu \to 0$, which
must be the case since in this limit $\Gamma_+$ goes
over smoothly to the flat space matrix $\Gamma$ of \gs, which is
not diagonal.

In the opposite limit $\mu \to \infty$ or $\lambda' \to 0$, we
note that $\Gamma_0$ is of order $\mu$ while $H$ is of order
$\mu^{-2}$.
Furthermore $\Gamma_0$ has a power series expansion around
$\mu = \infty$ in which only odd powers of $1/\mu$ appear,
while $H$ has an expansion with two kinds of terms:
even powers of $1/\mu$ and non-perturbative
terms of order $e^{-2 \pi \mu y}$ and $e^{-2 \pi \mu (1-y)}$
which come from the function $F(z)$.
The last kind of terms correspond to field theory
effects of order $e^{-1/\sqrt{\lambda'}}$, which
are reminiscent of D-branes rather than instantons.

We refer to $H$ as the `non-analytic' part of $\Gamma_+$ for
two reasons.  First, it is directly responsible
for the half-integer powers of $\lambda'$ and non-perturbative
$e^{-1/\sqrt{\lambda'}}$ effects.
Secondly, it is shown in appendix B how these terms arise
from a certain branch cut in the complex plane which was
missed in the analytic continuation argument
of \ChuPD.
We will see however that
$H$ also contributes
to integer powers of $\lambda'$ in  observables.

Having now an explicit expression for the elements of
the matrix $\Gamma_+$,
the next step is to find $\Gamma_+^{-1}$.
Since $\Gamma_0$ is easy to invert and
is larger than $H$ by a factor of $\mu^3$ for large $\mu$,
it seems sensible to employ the expansion
\eqn\expansion{  
\Gamma_+^{-1} = (\Gamma_0 - H)^{-1}
= \Gamma_0^{-1} + \Gamma_0^{-1} H \Gamma_0^{-1}
+ \Gamma_0^{-1} H \Gamma_0^{-1} H \Gamma_0^{-1} + \cdots.
}
In order to establish the validity of this expansion,
two issues must be addressed:  the first is whether each term
on the right-hand side is finite, and the second is whether the sum
of all of the terms converges.

Naive counting of $\mu$'s suggests that each
term in \expansion\ is suppressed relative
to the previous term by a factor of $\mu^{-3}$.
However, the matrix product in $H \Gamma_0^{-1} H$ involves
a sum of the form
\eqn\sone{
\sum_{p=1}^\infty
{\sin^2(\pi p y) \over \sqrt{p^2 + \mu^2}}
{p^2 \over (p^2 + \mu^2 x^2)(p^2 + \mu^2 z^2)}.
}
We evaluate this sum in appendix B and find that
it behaves for large $\mu$ like $\mu^{-2}$
rather than the naive $\mu^{-5}$.  This `renormalization'
by $\mu^3$ is a direct consequence of the large
$p$ behavior of \sone, which would
equal $\mu^{-5}$ times a cubically divergent sum if one
tried to take $\mu \to \infty$ before evaluating the sum.

So the good news is that in the expansion \expansion, each term
on the right-hand side exists (indeed we present
an explicit formula for the $k$-th term in appendix B),
but the bad news is
that all of the terms (except the first) are of order $\mu^{-4}$!
Therefore it is not clear that the sum of these terms
converges, although
this can still be the case if the $k$-th term is suppressed
by a coefficient which decreases sufficiently rapidly
with $k$.
While we have not
been able to prove convergence, numerical evidence
suggests that the expansion \expansion\ is
indeed sensible and converges rapidly
to $\Gamma^{-1}$.

To summarize, we have shown that for large $\mu$,
\eqn\gpresult{
\mu \left[ \Gamma_+^{-1} \right]_{mn} = \left[ {m \over 2} - {m^3 \over
4}\lambda' + {\cal{O}}(\lambda'^2) \right] \delta_{mn} + \lambda'^{3/2} R_{mn}
+ {\cal{O}}(\lambda'^{5/2}),
}
where the term in brackets is the expansion of $\Gamma_0^{-1}$
and $R_{mn}$ is nonzero and nondiagonal but has eluded
explicit evaluation since it requires summing an infinite
number of terms in \expansion.
This result highlights the fact that \expansion, while a true formula,
is not very useful for studying the small-$\lambda'$ expansion.
Hopefully a more clever method of inverting $\Gamma_+$ can be found.

\subsec{Some Neumann Matrix Elements}

In some Neumann matrix elements $[\Gamma_+^{-1}]_{mn}$ appears
on its own, but in others it must be multiplied on the left
and/or right by certain $\mu$-independent matrices
or vectors (see appendix A).
In this subsection we show that these summations renormalize
the contribution of the non-analytic terms $H$ by additional powers
of $\mu$, allowing them to contribute at order $\lambda'$ or
even $\sqrt{\lambda'}$ to the Neumann matrix elements.

The simplest Neumann matrix is\foot{All expressions
in this subsection are valid for positive indices $m,n$.
These are sufficient to determine the elements with negative indices via
fairly
simple relations \mntwo.}
\eqn\nthree{
\overline{N}^{(33)}_{mn} = \delta_{mn} -2{ (m^2 + \mu^2)^{1/4}   
(n^2+ \mu^2)^{1/4} \over \sqrt{mn}} \left[ \Gamma_+^{-1} \right]_{mn}.
}
Using \gpresult\ we see immediately that for large $\mu$,
\eqn\aaa{
\overline{N}^{(33)}_{mn} =
- {2 \over \sqrt{m n}} \lambda'^{3/2} R_{mn} +
{3 \over 8} n^4 \delta_{mn}
\lambda'^2 + \cdots,
}
which demonstrates the existence of half-integer powers of $\lambda'$
in string theory observables.

Next consider the Neumann coefficient
$\overline{N}^{(13)}_{0m}$, which at large $\mu$ involves
$\mu [ \Gamma_+^{-1} B]_m$,
where the vector $B$ is defined in appendix A.
Using the expansion \expansion, we expect
\eqn\aaa{
\Gamma_+^{-1} B = \Gamma_0^{-1} B + \Gamma_0^{-1} H \Gamma_0^{-1} B
+ \cdots.
}
Our counting from the previous subsection suggests that
the first term is ${\cal{O}}(\mu^{-1})$, while the second
and all higher terms are ${\cal{O}}(\mu^{-4})$.  However, the
vector product in $H \Gamma_0^{-1} B$ involves a sum of the form
\eqn\sumtwo{
{1 \over \mu}
\sum_{p=1}^\infty {\sin^2(\pi p y) \over \sqrt{p^2 + \mu^2}}
{1 \over p^2 + \mu^2 x^2},
}
which is ${\cal{O}}(\mu^{-3})$ rather than the naive ${\cal{O}}(\mu^{-4})$
for large $\mu$.
This renormalization by one power of $\mu$ is again the direct
result of the large $p$ behavior of \sumtwo, which would be linearly
divergent if one tried to first set $\mu=\infty$ and then perform the sum.
In appendix B we show that
\eqn\aaa{
\left[\Gamma_0^{-1}  H \Gamma_0^{-1} B\right]_m =
{1 \over 2 \pi^2 \mu^3} m^3 B_m + {\cal{O}}(\mu^{-5}).
}
The complete answer therefore has the form  
\eqn\cool{
\mu [\Gamma_+^{-1} B]_m = {m \over 2 } B_m + m^3 \lambda'
\left[ - {1 \over 4} + {1 \over 2 \pi^2} + \cdots \right] B_m
+ {\cal{O}}(\lambda'^2).
}
Calculating the exact coefficient in brackets would require
summing up the infinite number of terms on the right-hand side of
\expansion, which we have not been able to do, but numerical
evidence suggests that the quantity converges rapidly (to
$-{1 \over 4} + x$, where $x \approx {1 \over 16}$).\foot{This
introduces an apparent disagreement with the
field theory calculation of \ChuPD, since we find
that the correction
factor in
(56) and (58) should be
$1 -(\ha - 4 x) \lambda' n^2$ instead
of $1  -\ha \lambda' n^2$.  However, in the string
field theory calculation one also needs to include the prefactor
which may further modify the ${\cal{O}}(\lambda')$ correction.}
Although \cool\ shows that no half-integer powers of $\lambda'$
enter in the Neumann coefficients $\overline{N}^{(13)}_{0m}$, it
is remarkable that the coefficient of the ${\cal{O}}(\lambda')$
term receives a finite renormalization due to the
non-analytic contribution from $H$. 

A similar analysis holds for the Neumann coefficients
$\overline{N}^{(23)}_{0m}$, as well as
$\overline{N}^{(r3)}_{mn}$ for $r \in \{1,2\}$, 
although the latter involve a sum of the form
\eqn\aaa{
{1 \over \mu}
\sum_{p=1}^\infty {\sin^2(\pi p y) \over
p^2 - m^2/y^2} {1 \over \sqrt{p^2 + \mu^2}}
{p^2 \over p^2 + \mu^2 x^2},
}
rather than \sumtwo.  Like \sumtwo, this sum
behaves as ${\cal{O}}(\mu^{-3})$ for large $\mu$.
Therefore these Neumann matrix elements have no half-integer
powers, but it seems difficult to calculate
explicitly even the ${\cal{O}}(\lambda')$ term since all
of the terms in \expansion\ contribute, just as in \cool.

Finally, we remark that the remaining Neumann coefficients
involve $\Gamma_+^{-1}$ multiplied both on the left and on
the right.  For example, for $r,s \in \{1,2\}$, $\overline{N}^{(rs)}_{mn}$
involves $\mu [A^{(r) {\rm T}} \Gamma_+^{-1} A^{(s)}]_{mn}$, while 
$\overline{N}^{(rs)}_{0m}$ involves $\mu [A^{(r) {\rm T}} \Gamma^{-1} B]_m$.
In these cases there are two summations which each provide
an extra factor of $\mu$, so that these Neumann coefficients
have contributions starting at ${\cal{O}}(\sqrt{\lambda'})$.

\newsec{Conclusion}

In this paper we have studied the Neumann coefficients of the
three-string vertex in the pp-wave background.
These matrices are highly nontrivial functions of $\lambda'$
which smoothly interpolate between the weak and strong effective
coupling regimes of the SYM gauge theory and potentially
encode a wealth of information about non-BPS observables
in the field theory.
We have shown that these coefficients
contain half-integer powers of $\lambda'$ in the weak effective coupling
expansion. Recall, however, that the plane wave
limit is carried out at large `t Hooft coupling
$\lambda$. Therefore, there are
two possibilities. The first one is that
$(\lambda')^{n/2}$ may be replaced literally  by
$\lambda^{n/2}/J^n$, so that we find fractional powers of
$\lambda$ at weak coupling, as in the free energy \weak.
The second possibility is that $(\lambda')^{n/2}$
should be interpreted as $g(\lambda)/J^n$ where $g(\lambda)$
has a weak coupling expansion in integer powers of $\lambda$
but approaches $\lambda^{n/2}$ for large $\lambda$.
% Unlike \weak, the obstacles to calculating
%these coefficients for all $\lambda'$ are technical rather
%than conceptual.
It would be very desirable to
decide which of the two possibilities is correct.

We also remark that the precise relation between
the Neumann coefficients
and gauge theory three-point functions
is not well-understood at finite coupling.
This is both because the dictionary
between pp-wave string theory and SYM theory is not
precisely known
away from $\lambda'= 0$ (see \refs{\seven,\Herman}),
and because we have not included the prefactor of the
cubic string interaction \mntwo\ in our analysis,
although we do not expect the latter to change our conclusions
qualitatively.
Finally, the dictionary is also complicated by
mixing between single- and multi-trace operators.

\bigskip
\bigskip
\bigskip
\bigskip

\noindent
{\bf Acknowledgements}
\bigskip
We thank D. Freedman, N. Itzhaki and S. Minwalla for helpful discussions.
The work of I.~R.~K. was supported in part by the NSF
grant PHY-9802484.  M.S. and A.V. are supported in part by
DOE grants DE-FG02-91ER40671 and DE-FG02-91ER40654
respectively.

\appendix{A}{Some Matrices}

We start by defining for $m,n>0$ the matrices
\eqn\adef{\eqalign{
A^{(1)}_{mn} &= (-1)^{m+n+1} {2 \sqrt{m n} \over \pi} 
{y \sin(\pi m y)\over n^2-m^2 y^2},
\cr
A^{(2)}_{mn}
&=(-1)^{m} {2 \sqrt{m n} \over \pi} 
{(1-y) \sin(\pi m y) \over n^2-m^2 (1 -y)^2},
\cr
A^{(3)}_{mn} &= \delta_{mn},\cr
C_{mn} &= m \delta_{mn},\cr
C^{(1)}_{mn} &= \delta_{mn} \sqrt{m^2 + \mu^2 y^2},\cr
C^{(2)}_{mn} &= \delta_{mn} \sqrt{m^2 + \mu^2 (1-y)^2},\cr
C^{(3)}_{mn} &= \delta_{mn} \sqrt{m^2 + \mu^2}
}}
and the vector
\eqn\bdef{
B_m ={2 \over \pi y(1-y) \alpha' p^+} {(-1)^{m+1} \sin(\pi m y) \over
m^{3/2}}.
}
Note that $\mu = 1/\sqrt{\lambda'}$ stands for what was called
$\mu |\alpha_{(3)}|$ in \mn, $p^+ = p^+_{(3)}$ is the
momentum of the big string, and $y=p^+_{(1)}/p^+$ is the fraction
of $p^+$ carried by little string number 1.

The matrix $\Gamma_+$ whose inverse appears in the Neumann coefficients
for positive $m,n$ is given by \mntwo
\eqn\gpdef{
\Gamma_+ = \sum_{r=1}^3 A^{(r)} C_{(r)} C^{-1} A^{(r) {\rm T}} + \ha
\mu y (1-y) (\alpha' p^+)^2 B B^{\rm T}.
}
It is manifest that $\Gamma_+$ goes over smoothly to the matrix
$\Gamma$ of \gs\ as $\mu \to 0$.
The Neumann matrices are then given for $m,n>0$ by
\eqn\neumann{
\overline{N}^{(rs)}_{mn}
= \delta^{rs} \delta_{mn} - 2 \left[ C_{(r)}^{1/2} C^{-1/2} A^{(r) {\rm T}}
\Gamma_+^{-1} A^{(s)} C^{-1/2} C_{(s)}^{1/2} \right]_{mn}.
}

\appendix{B}{Some Sums and Integrals}

Let us first calculate $\Gamma_+.$  From
\gpdef\ and the definitions
\adef\ 
it is easy to see that we need to evaluate sums of the
form
\eqn\sone{
\sum_{p=1}^\infty f(p), \qquad f(z) = {\sqrt{z^2 + \mu^2 y^2}
\over (z^2 - m^2 y^2) (z^2 - n^2 y^2)},
}
(and the same with $y \to 1-y$)
for positive integers $m,n$ and $0<y<1$.
To this end consider the integral
\eqn\aaa{
I_C = 
\oint_{C} {dz \over 2 \pi i} f(z) \pi \cot(\pi z)
}
for the various contours shown in Fig.~1.

\fig{The analytic structure
of the function $f(z) \pi \cot(\pi z)$ for $f(z)$ given in \sone.
The poles lie at all integer $z$, with four additional
poles on the real axis at $z = \pm m y, \pm n y$.
The branch cuts on the imaginary axis start
at $z = \pm i \mu$.
The top, bottom and central contours
correspond to
$I_t$, $I_b$ and $I_c$
respectively.}
{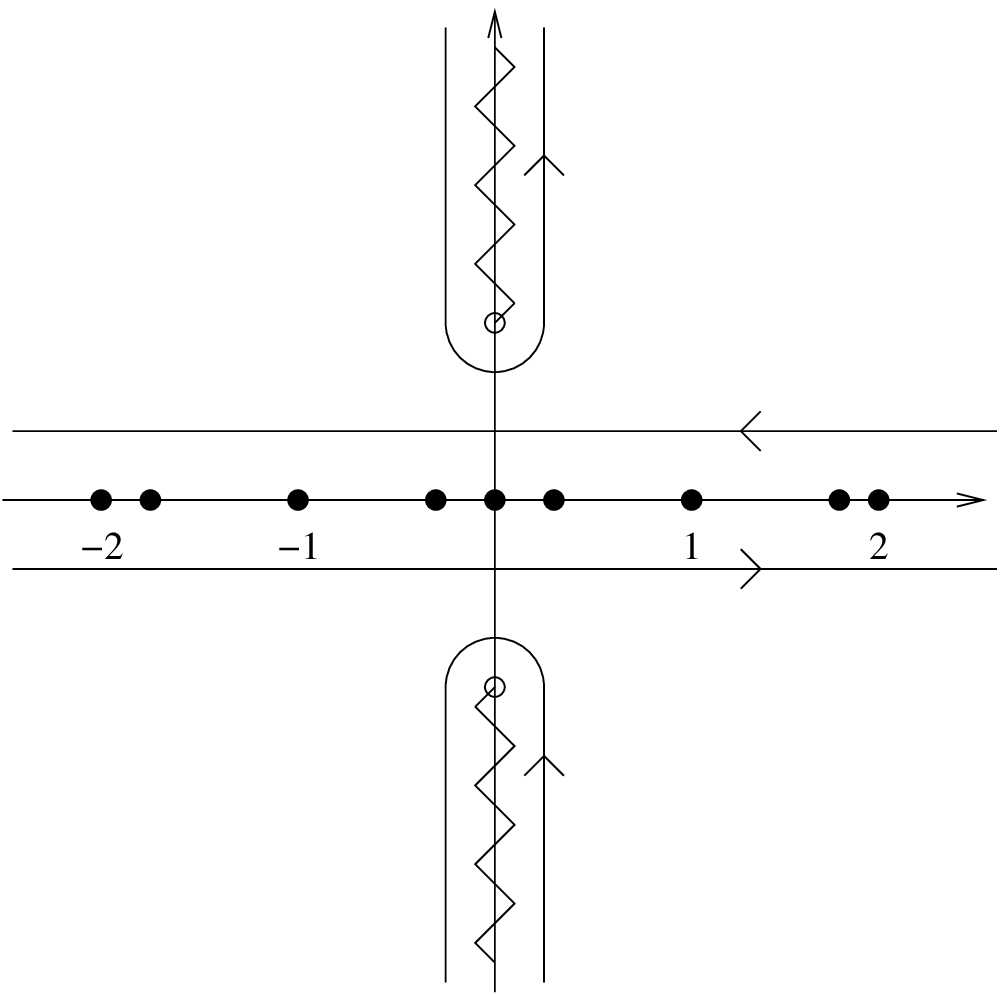}{2.0in}

It is easy to evaluate
\eqn\aaa{\eqalign{
I_c &={\mu \over
y^3 m^2 n^2} + {\pi \over y^2} 
\left[ {\cot(\pi m y) \over m(m^2-n^2)}
\sqrt{m^2  + \mu^2 }
+ (m \leftrightarrow n)\right] + 2 \sum_{p=1}^\infty f(p),\cr
I_t &= I_b
= {1 \over \mu^2 y^2} \int_1^\infty dx {\sqrt{x^2 - 1} \over
(x^2 + m^2/\mu^2)(x^2 + n^2/\mu^2)} \coth(\pi \mu y x).
}}
Now, since
\eqn\aaa{
I_t + I_b + I_c = 0,
}
we conclude that
\eqn\sumone{
\sum_{p=1}^\infty f(p) = - {\mu \over 2 y^3 m^2 n^2}
- {\pi \over 2 y^2} \left[
{\cot(\pi m y) \over m (m^2-n^2)} \sqrt{m^2+ \mu^2}
+ (m \leftrightarrow n)\right] - I_t.
}
Note that for very large $\mu$ we can set $F(z) = 1$
and evaluate the integral $I_t$, obtaining
\eqn\aaa{
I_t = {1 \over y^2} {1 \over (m^2 - n^2)}  
\left[ {\sqrt{m^2 + \mu^2} \over m} {\rm arcsinh} (m/\mu) -
{\sqrt{n^2 + \mu^2} \over n} {\rm arcsinh} (n/\mu)\right].
}
This is valid up to corrections of order
$e^{-2 \pi \mu y}$ and $e^{-2 \pi \mu (1-y)}$.

Using the sum \sumone\ and the definitions in appendix A,
it takes only a little algebra to show that the contribution
from the first two terms in \sumone\ is such that the $r=1,2$
terms in \gpdef\ cancel the $B B^{\rm T}$ term, leaving
only a diagonal piece $\ha \Gamma_0$.  The other $\ha \Gamma_0$
comes from $r=3$ in \gpdef.
The net result is that omitting $I_t$ in \sumone\ would
lead one to the conclusion that $\Gamma_+ = \Gamma_0$, as
in the analytic continuation argument of \ChuPD.
Instead, we find that the branch cut terms $I_t$
precisely account for the matrix $H$ as written
in \hdef\ after summing over $r=1,2$ in \gpdef.

Let us list some other useful sums which
can be derived using similar techniques.
For $v>1$ we have
\eqn\sumfive{
\sum_{n=1}^\infty {\sin^2(\pi n y) \over n^2 + \mu^2 v^2}
{1 \over \sqrt{n^2 + \mu^2}}
=-{1 \over 2 \mu^2}
P \int_1^\infty {dz \over \sqrt{z^2 - 1}}
{1 \over z^2 - v^2}
{1 \over F(z)},
}
where the symbol $P$ stands for the principal value of the integral.
For large $\mu$ we can set $F(z) = 1$ and evaluate the integral,
giving
\eqn\sumfour{
\sum_{n=1}^\infty {\sin^2(\pi n y) \over
n^2 + \mu^2 v^2} {1 \over \sqrt{n^2 + \mu^2}} = {1 \over 2 \mu^2}
{{\rm arccosh}(v) \over v \sqrt{v^2-1}},
}
up to exponential corrections.
A variant of \sumfive\ which we will need is
\eqn\aaa{
\sum_{n=1}^\infty {\sin^2 (n \pi y) \over
n^2 - m^2/y^2} {1 \over \sqrt{n^2 + \mu^2}}=
-{1 \over 2 \mu^2} \int_1^\infty {dz \over \sqrt{z^2 - 1}}
{1 \over z^2 + m^2/(\mu^2 y^2)}
{1 \over F(z)},
}
where $m$ is an integer.

Next we study the $k$-th
term in the expansion
\expansion.  Using the integral representation \hdef\ for $H$,
we find that
the matrix multiplication $H \Gamma_0^{-1} H$ involves
a sum of the form
\eqn\pdef{\eqalign{
P(x_1, x_2) &\equiv \ha \sum_{p=1}^\infty
{\sin^2(\pi p y) \over \sqrt{p^2 + \mu^2}}
{p^2 \over (x_1^2 + p^2/\mu^2) (x_2^2 + p^2/\mu^2)}
\cr
&= - {\mu^2 \over 4}
P \int_1^\infty {dz \over \sqrt{z^2 - 1}} {1 \over F(z)}
{z^2 \over (z^2-x_1^2) (z^2 - x_2^2)}.
}}
Using this definition it is straightforward to
derive the explicit though complicated formula
\eqn\cooltwo{
\eqalign{
&[H \Gamma_0^{-1} H \Gamma_0^{-1} \cdots H]_{mn}
= \left( {8 \over \mu^2 \pi^2} \right)^k (-1)^{m+n} \sqrt{m n}
\sin(\pi m y) \sin(\pi n y)\cr
&\qquad
\qquad\qquad\times \left[
\prod_{i=1}^k \int_1^\infty dx_i \sqrt{x_i^2 - 1} F(x_i)
\right]
{ P(x_1,x_2) \times \cdots \times P(x_{k-1},x_k) \over
(x_1^2 + m^2/\mu^2)(x_k^2 + n^2/\mu^2)},
}}
where $k$ is the number of times $H$ appears on the left.
Since each of the $k-1$ `propagators' $P$ has an explicit factor
of $\mu^2$ from the result \pdef, we see that \cooltwo\ is
${\cal{O}}(\mu^{-2})$ for any $k$.
This establishes the claim that all of the terms
in \expansion\ except for the first are ${\cal{O}}(\mu^{-4})$ for large
$\mu$.
Note that for large $\mu$ we can set $F(z) = 1$ in \pdef\ to obtain
\eqn\pdef{
P(x_1,x_2) = {\mu^2 \over 4}
\left[
{x_1 \over x_1^2 - x_2^2} { {\rm arccosh}(x_1)
\over \sqrt{x_1^2 - 1}} +
(x_1 \leftrightarrow x_2)\right],
}
up to exponential corrections.
Nevertheless we have not been able to evaluate the iterated
integrals in \cooltwo\ in a closed form.

Let us conclude
by calculating the second term in brackets in \cool.
We have
\eqn\aaa{\eqalign{
\left[H \Gamma_0^{-1} B\right]_m &= \ha \left[ H C C_{(3)}^{-1} B\right]_m\cr
&= \ha \sum_{n=1}^\infty {8 \over \mu^2 \pi^2} (-1)^{m+n}
\sqrt{m n} \sin(\pi m y) \sin(\pi n y)
\cr
&\qquad\times \int_1^\infty dz { F(z) \sqrt{z^2 - 1} \over
(z^2 + m^2/\mu^2) (z^2 + n^2/\mu^2)}
\cr
&\qquad\times {n \over \sqrt{n^2 + \mu^2}}
\times {2 \over \pi y(1-y) p^+}
(-1)^{n+1} n^{-3/2} \sin(\pi n y).
}}
The sum over $n$ can be evaluated for large $\mu$ using
\sumfour.
The remaining integral over $z$ is then of the form
\eqn\aaa{
\int_1^\infty dz  {{\rm arccosh}(z)
\over z^3}=
\ha.
}
Putting everything together, we find
for large $\mu$
\eqn\aaa{
[ \Gamma_0^{-1} H \Gamma_0^{-1} B]_m = 
{1 \over 2 \pi^2 \mu^3} m^3 B_m.
}

\listrefs
\end